%% file: fairtor.tex
\documentclass{sig-alternate}
\usepackage[square,numbers]{natbib}
\usepackage[latin1]{inputenc}
\usepackage[T1]{fontenc}
\usepackage{url}
\usepackage{color}
\usepackage[usenames,dvipsnames,table]{xcolor}
\usepackage{epsfig}
\usepackage[normalem]{ulem}
\usepackage{listings}
\usepackage{tikz}
\usepackage{float}
\usepackage{hyperref}
\usepackage[hyphenbreaks]{breakurl}
\usepackage{wrapfig}
\usepackage{multirow}
\usepackage{graphicx}
\usepackage{caption}
\usepackage{subcaption}
\usepackage{protocolj}
\usepackage{mdframed}

\makeatletter
\def\@copyrightspace{\relax}
\makeatother

\usetikzlibrary{decorations.pathmorphing}
\usetikzlibrary{arrows,shapes,trees,shadows}
\usetikzlibrary{positioning}
\usetikzlibrary{backgrounds}

\title{Fair anonymity for the Tor network}

\numberofauthors{3}

\author{
  \alignauthor
  Jesus Diaz\\
  \affaddr{Escuela Politecnica Superior}\\
  \affaddr{Universidad Autonoma de Madrid}\\
  \email{j.diaz@uam.es}
  \alignauthor
  David Arroyo\\
  \affaddr{Escuela Politecnica Superior}\\
  \affaddr{Universidad Autonoma de Madrid}\\
  \email{david.arroyo@uam.es}
  \alignauthor
  Francisco B. Rodriguez\\
  \affaddr{Escuela Politecnica Superior}\\
  \affaddr{Universidad Autonoma de Madrid}\\
  \email{f.rodriguez@uam.es}
}

\begin{document}
\maketitle

\def\Ui{$\texttt{U}_i$}
\def\Uj{$\texttt{U}_j$}
\def\EN{\texttt{EN}}
\def\EX{\texttt{EX}}
\def\com{\texttt{com}}
\def\fcom{{\sf Com}}
\def\ProveZK{{\tt ProveZK}}
\def\VerifyZK{{\tt VerifyZK}}

\begin{abstract}

  Current anonymizing networks have become an important tool for guaranteeing
  users' privacy. However, these
  platforms can be used to perform illegitimate actions, which sometimes makes
  service providers see traffic coming from these networks as a probable threat.
  In order to solve this problem, we propose to add support for \emph{fairness}
  mechanisms to the Tor network. Specifically, by introducing a slight 
  modification to the key negotiation process with the entry and exit nodes, in
  the shape of group signatures. By means of these signatures, we set up an access
  control method to prevent misbehaving users to make use of the Tor network. Additionally,
  we establish a predefined method for denouncing illegitimate actions, which impedes the
  application of the proposed fairness mechanisms as a threat eroding users' privacy.
  As a direct consequence, traffic coming from Tor would be considered less suspicious 
  by service providers.
\end{abstract}

\section{Introduction}
\label{sec:introduction}

Privacy has become a major concern for Internet users. One main
approach for enabling privacy is the introduction of anonymizing techniques, and
Tor \cite{dms04} is probably the most popular and widely used anonymizing network.

Tor anonymizes communications by avoiding origin and recipient to be linked.
Moreover, even the recipient cannot learn the IP address of the originator
by analyzing the received packets. This is achieved by re-routing the data through 
several intermediaries, the Onion Routers, and adding an extra layer of encryption
with each one. Nevertheless, this also reduces the protection available for the 
addressee, since it cannot \emph{denounce} the originator or even block him. This is 
certainly a factor hindering any wide acceptance of anonymizing networks. Moreover, 
it causes users acting legitimately to be affected by the illegitimate actions of others.
For instance, in some situations legitimate users cannot access a site through Tor, 
because that site directly bans Tor-originated traffic.

In this work we use group signatures to extend the functionality of Tor's entry 
and exit nodes in order to enable the tracing and blocking of misbehaving users.
This being the case, we implement an access control mechanism for Tor which does
not deteriorate the normal use of the Tor network by users acting legitimately. As 
a consequence of this \emph{fairness}\footnote{According to \cite{kty04}, with 
  \emph{fairness} we refer to the capability of taking any measure that could prevent 
  anonymity misuses.} 
mechanism, service providers would probably increase their trust in Tor, since 
illegitimate actions coming from Tor would presumably be reduced. 

Section \ref{sec:mistrust} summarizes some data evidencing that irrevocable
anonymity and unlinkability may be seen as a threat, or something undesirable,
in some situations, and in Section \ref{sec:related}, we summarize some 
related work. We then describe our approach for incorporating fairness into 
the Tor network in Section \ref{sec:incorporatingfairnesstor}. Section 
\ref{sec:issues} points out to some open issues that are important to take
into account for realizing our proposal. Finally, Section \ref{sec:conclusion} 
concludes with a summary of our proposal. Throughout this short paper we assume a basic 
knowledge of Tor essentials. For a more detailed description of Tor, we refer 
to \cite{dms04} or the Tor project website\footnote{\url{https://www.torproject.org}}.

\section{Mistrust in complete anonymity}
\label{sec:mistrust}

Despite the fact that an obvious use of communications anonymity is to protect users' 
privacy, it can also be misused. For instance, in \cite{mbgks08} the authors run 
a Tor exit node for gathering statistical evidence for a further 
analysis of Tor's traffic. They conclude stating that it is not uncommon 
to see \emph{``hacking attempts, allegations of copyright infringements, and bot 
  network control channels''} routed through Tor. Another type of behavior that 
may be considered as undesirable is the use of the
Tor platform for purposes other than the originally intended ones, even though
they might not be against any ethical rule. This is reported in \cite{cmk10} 
as a result of analyzing the traffic going through several Tor exit nodes, and 
concluding that an important share of the packets being routed through Tor 
corresponds to BitTorrent traffic (roughly the 25\%). This reflects that in the case of willing 
to limit this use of Tor, little more than actually blocking all the BitTorrent 
traffic could be done. This would be unfair to users routing BitTorrent traffic 
through Tor but consuming a moderate portion of the available bandwidth. Hence, a 
finer control on this type of ``misbehavior'' is also desirable.
More evidence supporting the claim that the anonymity provided by Tor can be 
(and is being) misused is the fact that there already exist services, like 
BlockScript\footnote{\url{http://www.blockscript.com/features.php}.}, which 
include explicit functionality for blocking traffic, including data coming from
Tor and other anonymizing networks (BlockScript, offers a commercial 
service for blocking ``unwanted'' traffic and sells the raw blacklist data for 
\$12,000 per year). An interesting discussion about the necessity of accountability
in anonymous communication systems is done in \cite{dp07}. Finally, the risk
of failure due to websites blocking Tor has also been considered by the Tor
staff too in a recent post in the Tor Project blog: \emph{``A call to arms: Helping 
  Internet services accept anonymous users''}%
\footnote{\url{https://blog.torproject.org/blog/call-arms-helping-internet-services-accept-anonymous-users}.}.

Furthermore there is 
an abuse specific FAQ in the official Tor website\footnote{\url{https://www.torproject.org/docs/faq-abuse.html.en}.}
dealing with the subject.

\section{Related work}
\label{sec:related}

Several systems have been proposed to endow anonymizing platforms 
with fairness mechanisms. BLAC \cite{taks07} makes use of a specific group signature 
scheme in order to allow service providers to manage their own blacklists, following 
their own judgment when it comes to block users. Nymble \cite{tkcs11} creates a complex 
infrastructure in order to provide fairness by revoking the unlinkability of users 
who misbehave. In PEREA \cite{atk11}, the need of Trusted Third Parties is eliminated 
at the cost of creating a highly crafted infrastructure in order to allow users 
to be blocked.  EPID \cite{bl12} makes unlinkability revocation possible by either
using a member private key, a signature issued by the member to revoke, or by 
consulting the issuer of the member key. However, it requires the usage of Trusted 
Privacy Modules \cite{TPM03}, which we consider out of the scope of our proposal.

These systems also take into account that different misbehaviors are possible and,
in the case of PEREA, that each misbehavior should be assigned a different severity
(the PEREA-Naughtiness variant). Nevertheless, they would probably be too rigid 
for multi-purpose contexts where different legitimate uses and revocation needs 
are possible. For instance, Nymble just supports unlinkability revocation during
a predefined interval. BLAC is tied up to a specific group signature scheme, that 
might not be appropriate for some situations. PEREA limits the number 
of authentications that users may perform during a time span, blacklisting them
if they exceed that limit. Hence, any change on the implemented 
functionality that may be necessary for the adoption of these systems into a new 
context would probably require a great effort. 

Our proposal takes advantage of the wide variety of group signatures and the
standardization process that anonymous certificates based on them are being
subject to \cite{bcly07,dar14,abc4trust,isoiec200081}. These would allow a
high flexibility for modifying the desired functionality depending on the
context and, at the same time, the minimization of the deployment costs.

\section{Building blocks}
\label{sec:building}

We use the notation $\langle O_A, O_B \rangle \leftarrow Pro (I_C) \lbrack
A(I_A),B(I_B)\rbrack$ to describe a two-party process $Pro$ between parties 
$A$ and $B$, where $O_A$ (resp.  $O_B$) is the output to party $A$ (resp. $B$), 
$I_C$ is the common input, and $I_A$ (resp.  $I_B$) is $A$'s (resp. $B$'s)
private input; when party $B$ does not have output, we sometimes write 
$O_A \leftarrow Pro (I_C) \lbrack A(I_A),B(I_B)\rbrack$.
Single-party processes are denoted by $O \leftarrow Pro(I)$, with input
$I$ and output $O$. Also, for readability, we omit the publicly known keys
in the process calls. Finally, since we continuously deal with different 
types of signatures, we use the Greek letter $\sigma$ for denote a signature 
produced by some of the schemes described below (the specific type is clear
given the notation in the following subsections).

\subsection{Group signatures}
\label{ssec:groupsig}

Group signatures, first proposed by Chaum and Van Heyst \cite{ch91}, allow
a member of a group to issue a signature such that any possible verifier can 
check that it has been issued by a member of the group, without revealing
which specific member issued it. More advanced schemes have been proposed since 
then \cite{cl02,kty04,lpy12b}, improving the scalability and efficiency of group 
signatures, but also their functionality. Current schemes allow unlinkability 
revocation and anonymity revocation. Normally, these revocations can only be 
performed by suitable authorities, but there exist schemes that introduce 
interesting variants to this behavior, allowing unlinkability revocation only for 
users who have broken some rule (like exceeding the maximum number of messages to 
sign \cite{tfs04}). 

To summarize, a group of a group signature scheme is basically composed by:
a group manager who owns a secret group manager key $MK$ and publishes a 
group key $GK$; and a set of $N$ members each in possession of a member key
$mk_i$, for $1\le i \le N$. Overall, the main operations supported by a group 
signature scheme may be summarized as follows:

\begin{description}
\item[$MK, GK \leftarrow \texttt{GS.Setup}(1^k)$.] Run by the group manager, 
  creates the group key and the manager key.
\item[$mk_i \leftarrow \texttt{GS.Join}\lbrack U(secret), M(MK)\rbrack$.] 
  Executed jointly between a new user $U$ and the group manager $M$, allows 
  new users to join the group, obtaining a member key.
\item[$\sigma \leftarrow \texttt{GS.Sign}(msg,mk_i)$.] A user creates a group 
  signature over a message, using her member key.
\item[$b \leftarrow \texttt{GS.Verify}(\sigma,msg)$.] Allows anyone with the 
  group key to verify a group signature.
\item[$trapdoor \leftarrow \texttt{GS.Open}(\sigma,MK)$.] Run by the group manager, 
  allows him to obtain de-anonymize to some extent the issuer of a group signature,
  given the group signature itself, the manager key and, normally, some additional 
  private information. In some 
  schemes, the only possibility is to retrieve her real identity. In others, it is 
  possible to obtain a token which permits to link group signatures made by the 
  same user. For simplicity, in this work we encompass all variants that somehow 
  reduce anonymity under the term \texttt{Open}.
\item[$b \leftarrow \texttt{GS.Trace}(\sigma,MK)$.] Allows to verify if a given group 
  signature has been issued by some arbitrary member. In order to run this operation, 
  \texttt{GS.Open} needs to be executed before. Thus, note that there may be different 
  ways to achieve this, according to the variants of the \texttt{GS.Open} functionality.
\end{description}

However, this
can be considered a ``general working mode'' of a group signature scheme. For a 
good review of the main advancements in the field of group signatures, we refer
to \cite[Section 1.1]{lpy12b}.

\subsection{Blind signatures}
\label{ssec:blindsig}

\emph{Blind signatures}\index{Blind signatures} where introduced by Chaum 
in \cite{chaum82}. Basically, a blind signature scheme allows a user $U$ to obtain 
a signature from a signer $S$ over any arbitrary message $m$, but without $S$ 
learning anything about $m$. This idea has been used since then for creating 
diverse privacy respectful systems. However, since the 
signer does not learn any information about the message, systems based on them 
can easily be abused. For solving this issue, \emph{fair blind signatures}
were proposed in \cite{spc95}. In this variant, an authority has privileged 
information allowing the signer to link message and signature pairs. 
\emph{Restrictive blind signatures} \cite{brands93} allow issuing
blind signatures, but only choosing among messages that comply certain rules.
Finally, an also important alternative is given by partially blind signatures 
\cite{af96}. In a \emph{partially blind signature} the messages are divided in 
two parts: a common message to which the $S$ has complete access; and the 
blinded message, of which $S$ does not learn anything. Thus, the common message
may be employed to implement misuse prevention mechanisms. As always, several 
schemes have appeared improving the overall efficiency, reducing the size of 
the final signatures, or based on different number theory problems \cite{jl99,
  czms06,okamoto06,bfpv13}.

Although some of the variants of blind signatures would probably be useful for 
our proposal, we use the general definition of a blind signature for describing 
it. Thus, hereafter we will use the following notation when referring to the 
operations of a blind signature scheme:

\begin{description}
\item[$(pbk, sbk) \leftarrow \texttt{BS.Setup}(1^k)$.] Creates the signer's public $pbk$ 
  and private keys $sbk$ for issuing blind signatures.
\item[$(\beta,\pi) \leftarrow \texttt{BS.Blind}(msg,secret)$.] Using some random secret 
  value, the user creates a blinded version ($\beta$) of the message to be blindly 
  signed and a proof of correctness $\pi$.
\item[$\tilde \beta \leftarrow \texttt{BS.Sign}(\beta,sbk)$.] Upon receiving the 
  blinded messages, the signer runs any necessary verification and creates a blinded 
  signature using its private key.
\item[$\sigma \leftarrow \texttt{BS.Unblind}(\tilde \beta, secret)$.] The user receives 
  the blinded signature and unblinds it, using the secret value generated during the 
  blind process. The result of this operation is the final signature.
\item[$b \leftarrow \texttt{BS.Verify}(\sigma,msg)$.] Any entity runs this operation 
  to verify the signature.
\end{description}

\subsection{Blind group signatures}
\label{ssec:blingdroupsig}

A blind group signature scheme is just like a blind signature in which the signer
issues a group signature instead of a conventional signature. Therefore, 
each of the operations described in the previous sections may be independently 
applied in this schemes. However, for the sake of clarity, when referring to
this schemes, we will use the prefix \texttt{BGS} instead of \texttt{GS} or 
\texttt{BS}.

\subsection{Additional cryptographic primitives}
\label{ssec:additional}

Besides the primitives introduced above, we assume readers are familiar with 
public-key encryption, digital signature and commitment schemes, and zero-knowledge 
proofs of knowledge. We use $\com \leftarrow \fcom(m,r)$ to denote 
a commitment $\com$ to a message~$m$, where the sender uses uniform random 
coins~$r$; the sender can open the commitment by sending $(m, r)$ 
to the receiver. We use $\pi \leftarrow \texttt{ProveZK}(x,w)$ and
$\texttt{VerifyZK}(x, \pi)$ to refer to creating non-interactive proof
$\pi$ showing that the statement $x$ is in the language (which will be
determined by the context) with the witness $w$, and to verifying the
statement $x$ based on the proof $\pi$.

\section{Guidelines for incorporating\\ fairness into Tor}
\label{sec:incorporatingfairnesstor}

In order to endow Tor with fairness capabilities, the entry and exit nodes take a
central role, since they are the only nodes who learn the IP addresses of the user
entering the network and that of the final destination, respectively. Hence, their 
knowledge would be necessary to determine whether the IP trying to access the 
network has already been blocked, or to demonstrate that a given origin IP has 
accessed certain destination IP. However, when proposing modifications of those 
nodes we must avoid enabling attacks based on establishing
a connection between them. For that purpose, we take advantage of both the way the 
user negotiates keys with the Tor nodes, and the properties of group and blind 
signatures. 

Hereafter, we assume that a group has already been set up, and that there is a
suitable policy established for fairly managing revocation (see Section 
\ref{sec:issues}). %
Similarly, we assume that the blind signature scheme has also been set up.
Table \ref{tab:notation} summarizes the notation used throughout the rest paper, 
along with some notation inherited from the description of the Tor network \cite{dms04}
and the notation defined for group and blind signatures and the additional
cryptographic primitives in Section \ref{sec:building}.

\begin{table}[ht!]
  \centering
  \begin{scriptsize}
    \begin{tabular}{|c|l|}
      \hline
      $\{\cdot\}_K$ & Symmetric encryption with key $K$\\
      $\{\cdot\}_{PK_A}$ & Asymmetric encryption with A's public key.\\
      $\{\{\cdot\}\}$ & Layered encryptions following the Tor protocol.\\
      $H(\cdot)$ & Application of a cryptographic hash function.\\
      $g^{x_\cdot}$ & The user's Diffie-Hellman share.\\
      $g^{y_\cdot}$ & The Diffie-Hellman share corresponding to a Tor node.\\
      $hs_K$ & A transcription of the handshake for key $K$.\\
      $A|B$ & $A$ concatenated with $B$.\\
      $\sigma_1$ & Group signature of $g^{x_1}$ sent to entry node.\\
      $\sigma_2$ & Group signature of $g^{x_2}$ sent to exit node.\\
      $\beta$ & Blinded version of $\sigma_2$ \\
      $\tilde \beta$ & Blindly signed version of $\beta$ \\
      $\sigma_3$ & Blind signature of $\sigma_2$ \\
      \hline
    \end{tabular}
    \caption{Notation summary.\label{tab:notation}}
  \end{scriptsize}
\end{table}

Our approach works by introducing variations in the way a user negotiates
the symmetric keys with the entry and exit nodes. In short, we will require
the user to group-sign the message sent during negotiation with the entry 
and exit nodes. In addition, in order to prevent the user to employ one
identity for negotiating with the entry node, and a different one with the
exit node (see Section \ref{sec:issues}), the entry node has to blindly 
sign the message that the user will send to the exit node. The resulting 
modified handshake schemes (see \cite[p. 6]{dms04}) are shown below, where
\Ui~denotes any arbitrary user, \EN~denotes the entry node and \EX~the
exit node. During the handshake with \EN, \Ui~first group-signs $g^{x_1}$
and $g^{x_2}$, sends $g^{x_1}$ to \EN~and also requests \EN~to blindly 
sign a group signature of $g^{x_2}$. If all the operations succeed, \EN
accepts the connection.

\begin{center}
  \begin{minipage}[t]{\textwidth}
    \footnotesize
    \begin{tabbing}
      X\=X\=X\=\kill \textbf{Entry Node Handshake}: \\
      \>\Ui: $\sigma_1 \leftarrow$ \texttt{GS.Sign}$(g^{x_1}, mk_i)$ \\
      \>\Ui: $\sigma_2 \leftarrow$ \texttt{GS.Sign}$(g^{x_2}, mk_i)$ \\
      \>\Ui: $\com \leftarrow \fcom(\sigma_2,r_1)$ \\
      \>\Ui: $(\beta,\pi) \leftarrow$ \texttt{BGS.Blind}$(\com,r_2)$ \\
      \>\Ui: $\phi \leftarrow \ProveZK(x, w)$ where \\
      \>\>$x = (\beta, \pi, \sigma_1), w = (mk_i, r_1, r_2)$ such that: \\
      \>\>\>$\sigma_2 \leftarrow \texttt{GS.Sign}(g^{x_2}, mk_i)$, \\
      \>\>\>$(\beta, \pi) \leftarrow \texttt{BGS.Blind}(\fcom(\sigma_2,r_1),r_2)$ \\
      \>\Ui $\rightarrow$ \EN: $g^{x_1}, \sigma_1, \beta, \pi, \phi$ \\
      \>\EN: \VerifyZK($\beta, \pi, \phi, \sigma_1$) \\
      \>\EN: \texttt{GS.Verify}$(\sigma_1, g^{x_1})$ \\
      \>\EN: $\tilde{\beta} \leftarrow$ \texttt{BGS.Sign}$(\beta,sbk)$ \\
      \>\EN: $K_1 = g^{x_1y_1}$ \\
      \>\EN $\leftarrow$ \Ui: $g^{y_1},\tilde \beta, H(K_1|hs_{K_1})$ \\
      \>\Ui: $\sigma_3 \leftarrow$ \texttt{BGS.Unblind}$(\tilde{\beta},r_2)$ \\
      \>\Ui: $K_1 = g^{x_1y_1}$ \\
    \end{tabbing}
  \end{minipage}
\end{center}

When \Ui~initiates the handshake with \EX, she sends the group signature
on $g^{x_2}$ that was blindly signed by \EN, along with the blind signature 
itself. If all the verifications succeed, then \EX~accepts the connection.

\begin{center}
  \begin{minipage}[t]{\textwidth}
    \begin{tabbing}
      X\=X\=X\=\kill \textbf{Exit Node Handshake}: \\
      \>\Ui $\rightarrow$ \EX: $g^{x_2}, \sigma_2, \sigma_3$ \\
      \>\EX: \texttt{GS.Verify}$(\sigma_2, g^{x_2})$ \\
      \>\EX: \texttt{BGS.Verify}$(\sigma_3, \sigma_2)$ \\
      \>\EX: $K_2 = g^{x_2y_2}$ \\
      \>\EX $\rightarrow$ \Ui: $g^{y_2}, H(K_2|hs_{K_2})$ \\
      \>\Ui: $K_2 = g^{x_2y_2}$ \\
    \end{tabbing}
  \end{minipage}
\end{center}

It is important to note that the group signatures are encrypted using the public 
keys of either the entry or exit nodes. Hence, only the entry and exit nodes
learn them. Moreover, the group signature sent to the exit node is blindly signed
by the entry node. Thus, even if both nodes collude, they would not be able to 
determine by themselves that the group signatures they have received have been 
issued by the same user, due to the unlinkability property of the group signature 
scheme and the blindness property of the blind signature scheme. Moreover, since 
the group signature sent to the exit node has been blindly signed by the entry 
node, it is not possible for a user \Ui~to frame another user \Uj.

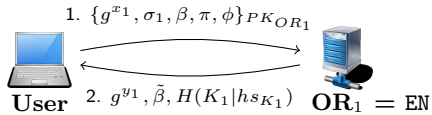
\begin{figure}[ht]
  \centering
  \input{fairtorhandshake1.tex}
  \caption{The user sends to the entry node a group signature of her
    share of the key, encrypted with the node's public key, and a
    blinded version of the group signature to be sent to the exit node.
    The entry node returns a blindly signed version of the latter.
    \label{fig:fairtorhandshake1}}
\end{figure}

\begin{figure}[ht]
  \centering
  \scalebox{0.9}{\input{fairtorhandshake2.tex}}
  \caption{The user sends to the exit node a group signature of her
    share of the key, encrypted with the node's public key and the
    blind signature issued by the entry node.
    \label{fig:fairtorhandshake2}}
\end{figure}
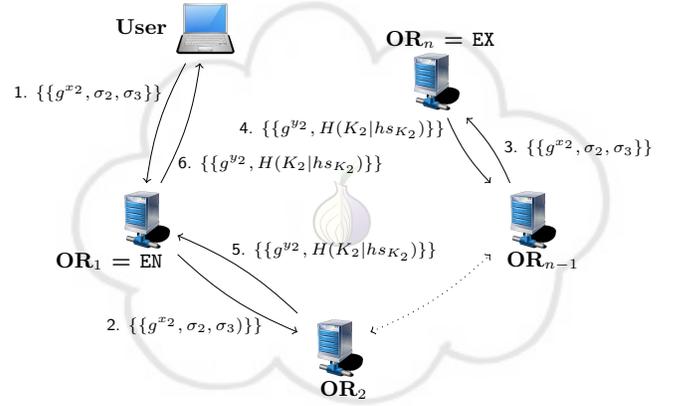

The modified key negotiation with the entry node is depicted in Fig. 
\ref{fig:fairtorhandshake1}, and the one corresponding to the exit node is depicted 
in Fig. \ref{fig:fairtorhandshake2}.

\subsection{How to block misbehaving users}
\label{ssec:howtoblock}

Let us assume that some user \Ui~ has been revoked due to some illegitimate behavior. 
When \Ui~ tries to establish a circuit,
he/she will need to perform a handshake with the chosen Tor entry node. Hence, upon 
receiving the first message with the group signature, the entry node will verify
the received group signature, checking whether or not the member who issued it 
has been revoked. Given that the member key of \Ui~ has been revoked, the
verification will fail, and the entry node will reject the connection. Note that
if the user has not been revoked, the privacy guarantees provided by Tor are
not diminished.

\subsection{How to denounce misbehaving users}
\label{ssec:howotodenounce}

In this case, we assume that \Ui~ has already established a circuit and she is 
communicating with some server $S$ (external to Tor). Also, let us suppose that 
eventually, \Ui~ performs some illegitimate action. When that happens, $S$ 
denounces this behavior following some predefined method. If deemed appropriate,
the group signature received by the exit node during the handshake may be used
to retrieve \Ui's identity, or to trace her. Specifically, the exit node provides
the following information:

\begin{itemize}
\item $\{msg\}_K$, where $msg$ is the message received and denounced by $S$, and
  $K$ is the symmetric key negotiated between \Ui~ and the exit node.
\item $(K = g^{x_2y_2},g^{x_2})$, where $g^{x_2}$ is \Ui's share of the handshake
  and $g^{y_2}$ is the share created by the exit node.
\item $\sigma_2$, i.e., a group signature of $g^{x_2}$ issued by \Ui.
\end{itemize}

In order to verify that the received denounce is valid, it is necessary to check
that the message received from $S$, $msg$, corresponds to the  encryption 
$\{msg\}_K$ received from the exit node. Also, $\sigma_2$ must be a valid group
signature over $g^{x_2}$. Finally, the exit node may be required to prove that
it knows the discrete logarithm $y_2$ of $g^{x_2y2}$ to the base $g^{x_2}$.
If these checks succeed, then the member with key $mk_i$ (\Ui) is responsible of 
$msg$. Hence, \Ui's key can be consequently revoked, and the circuit may be
closed by the exit node. Note that subsequent attempts made by \Ui~to establish 
a circuit would be blocked by the entry node, since the member key of \Ui~
has been revoked.

\subsection{Additional remarks}
\label{ssec:remarks}

A few remarks are worth to be made, concerning ``special'' situations and subjects 
that should be taken into account.

\begin{description}

\item[Tor bridges] Our proposal is directly extensible to support Tor bridges, 
  considering Tor bridges as the entry nodes to the Tor network.

\item[Leaky pipe topology] This approach allows a client to output some packets 
  through a node other than the negotiated exit. In order to adapt our approach 
  to this exception, the node leaking the packet should follow ad-hoc the procedure
  defined for exit nodes.

\item[Logging] Since hosts being accessed through these \emph{fair} Tor extension
  might want to issue denounces against traffic originated from Tor, logging the
  information necessary for solving disputes is required. Hence, the exit nodes
  need to keep the information specified in Section \ref{ssec:howotodenounce}
  (this also applies to the leaky pipe extension). An appropriate policy for
  logging should be established. 
  
\item[Denunciation time span] Considering that our proposal requires Tor exit
  nodes (and circumstantially other nodes) to log several pieces of information,
  it would comprise a serious scalability problem if this logging would be
  expected to last too much time. Hence, it seems appropriate to establish a
  predetermined time span for accepting denounces. Upon expiration of that time 
  span, all the logged information could be removed, and any subsequent denounce 
  related to that information rejected. This would require possible complainants
  to be aware of this time limitation.
\end{description}

\section{Open issues}
\label{sec:issues}

In the scheme given in Section \ref{sec:incorporatingfairnesstor} we just use 
the general definitions of the building blocks for defining our
system. The analysis of which specific variants should be employed is left as
future work. Note that this is a very delicate decision, since different options
offer different privacy properties. Moreover, we may even need different schemes 
depending on who issues the signatures (e.g. group signatures are issued both by 
users and entry points in our proposal). Thus, given the sensitivity of the 
information managed by Tor, this is an issue that needs to be studied in depth 
by itself. For that matter, the extensible group signatures library
\texttt{libgroupsig} \cite{dar15} may offer interesting features. In addition, concerning
the blind signatures, it would probably be necessary to use some of its variant
to prevent circumventing the controls explaining above. Namely, with the previous
bare scheme, a user could use the same blind signature indefinitely. This may
simply be solved by using partially blind signatures, and having the entry
node introduce a \emph{lifetime} value for the blind signature as common message.

Another important issues are determining when misbehaving users should be 
revoked, and by whom. The former question would probably depend on the websites
(or service) being accessed through Tor. For the latter, a probably good solution 
given Tor's infrastructure would be to apply threshold schemes to the revocation
procedures (see \cite{bcly08}), such that a majority of the authorities 
participating in the network consensus need to agree for revoking users.

Finally, note that Sybil attacks \cite{douceur02} are partly addressed by forcing users 
to use the same member key for the group signature sent to the entry node and for 
the group signature sent to the exit node (and having the latter to be blindly signed
by the entry node). However, some additional mechanism should be included for
preventing users from arbitrarily generating new member keys. Since asking users
to register may not be well received (it may be seem as a threat to anonymity),
requesting them to perform some proof of work \cite{dn92} during the generation 
of the member keys may be a good alternative.

\section{Conclusion}
\label{sec:conclusion}

In this work we have proposed an extension to the Tor network in order to endow
it with the functionality for preventing misbehaving users to access the network.
We expect such functionality to increase the trust of websites in Tor and thus
prevent them to block users coming from it. This extension follows
the design of Tor, and does not require any modification to its infrastructure.
It works by including two group signatures in the key negotiation processes with 
the entry and exit nodes (one for each) and having the entry node blindly sign
the group signature to be sent to the exit node. The group signature sent to the exit
node allows service providers to denounce illegitimate actions. Once the unlinkability
of a user has been revoked as a consequence of some illegitimate behavior, any entry
node would be able to block that specific user just by checking if it is included in
an (unlinkability) Revocation List.

\bibliographystyle{abbrv}
\bibliography{fairtor}

\end{document}

%% file: fairtorhandshake1.tex
\begin{tikzpicture}[scale=0.8,->]

  \node (client) at (0,1)  { \includegraphics[width=0.85cm]{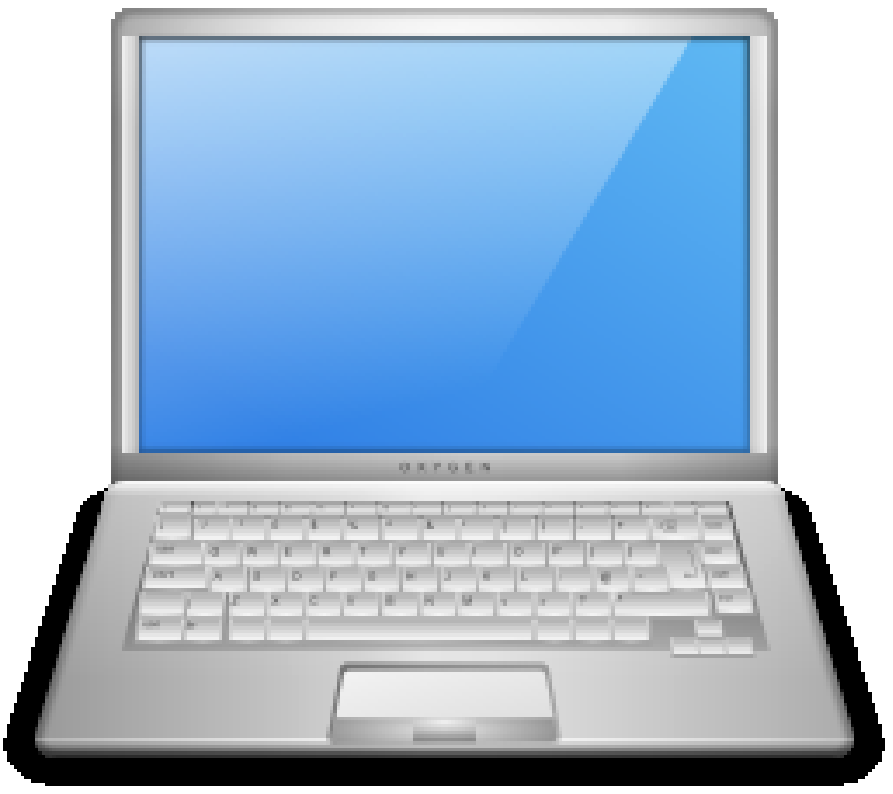} }; \node at (0,0.25) { \textbf{User} };
  \node (or1) at (5,1) { \includegraphics[width=0.85cm]{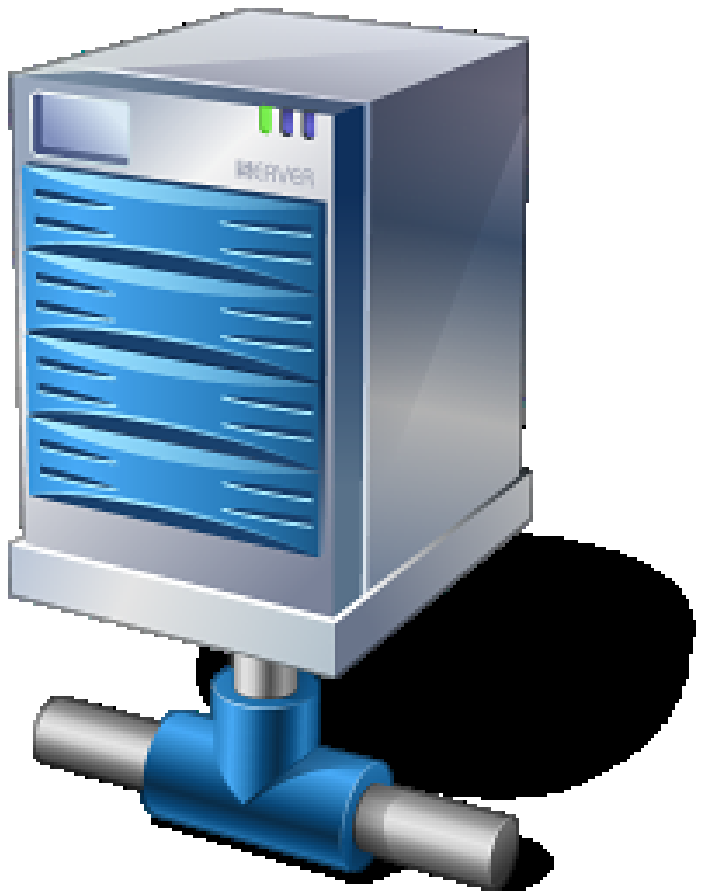} }; \node at (5.5,0.25) { \textbf{OR$_1$ = \EN} };  

  \path[every node/.style={font=\sffamily\small}]
  (client) edge [bend left=10] node[pos=0.50,above] {\scriptsize 1. $\lbrace g^{x_1}, \sigma_1, \beta, \pi, \phi \rbrace_{PK_{OR_1}}$} (or1)
  (or1) edge [bend left=10] node[pos=0.50,below] {\scriptsize 2. $g^{y_1}, \tilde \beta, H(K_1|hs_{K_1})$} (client);

\end{tikzpicture}

%% file: fairtorhandshake2.tex
\begin{tikzpicture}[scale=0.95,->]
  
  \node[opacity=0.1] at (3.15,3.25) { \includegraphics[width=8.5cm]{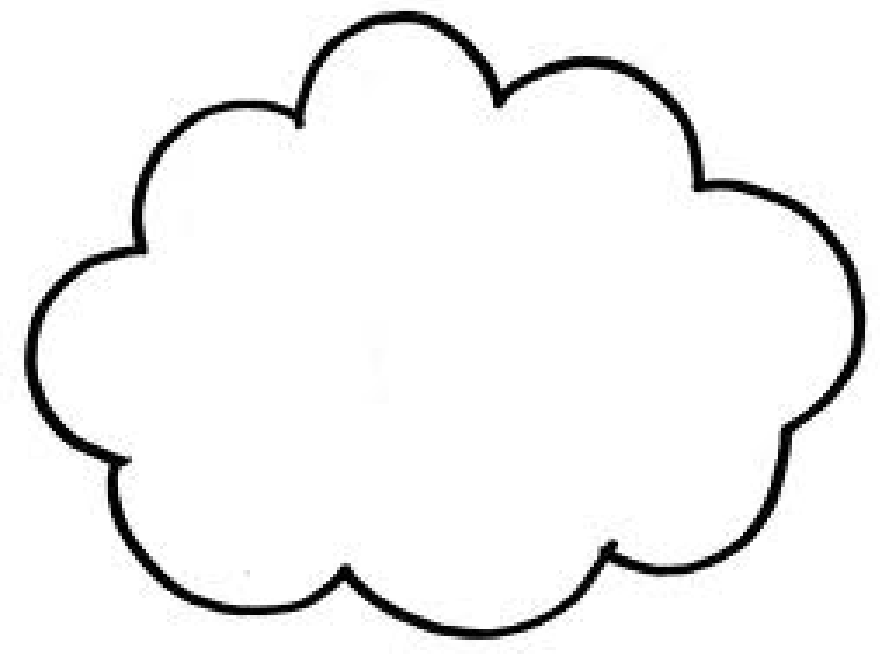} };
  \node[opacity=0.15] at (3.15,3.25) { \includegraphics[width=1cm]{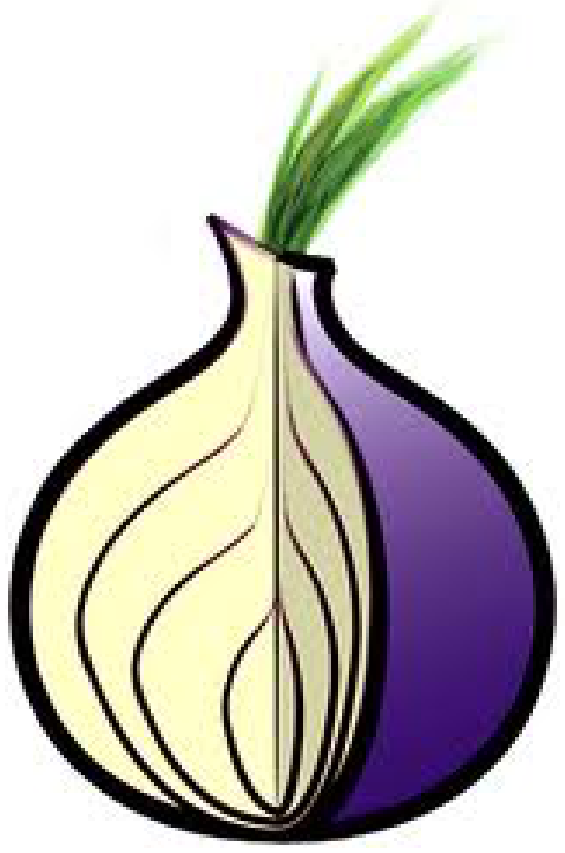} };
  \node (client) at (1,6)  { \includegraphics[width=0.85cm]{laptop.eps} }; \node at (0,6) { \textbf{User} };
  \node (or1) at (0,3) { \includegraphics[width=0.85cm]{server.eps} }; \node at (-0.5,2.35) { \textbf{OR$_1$ = \EN} };
  \node (or2) at (3,1) { \includegraphics[width=0.85cm]{server.eps} }; \node at (3.15,0.35) { \textbf{OR$_2$} };
  \node (or3) at (6,3) { \includegraphics[width=0.85cm]{server.eps} }; \node at (6.25,2.35) { \textbf{OR$_{n-1}$} };
  \node (or4) at (4.50,5.15) { \includegraphics[width=0.85cm]{server.eps} }; \node at (4.65,5.80) { \textbf{OR$_n$ = \EX} };
  
  \path[every node/.style={font=\sffamily\scriptsize}]
  (client) edge [bend right=10] node[pos=0.25,left,align=left] {\scriptsize 1. $\lbrace\lbrace g^{x_2},\sigma_2, \sigma_3 \rbrace\rbrace$} (or1)
  (or1) edge [bend right=10] node[pos=0.90,left=5pt] {\scriptsize 2. $\lbrace\lbrace g^{x_2}, \sigma_2, \sigma_3) \rbrace\rbrace$} (or2)
  (or2) edge [<->,bend right=10, dotted] node[pos=0.05,right=5pt] {} (or3)
  (or3) edge [bend right=10] node[pos=0.50,right=2pt] {\scriptsize 3. $\lbrace\lbrace g^{x_2}, \sigma_2, \sigma_3 \rbrace\rbrace$} (or4)
  (or4) edge [bend right=10] node[pos=0.15,left] {\scriptsize 4. $\lbrace\lbrace g^{y_2}, H(K_2|hs_{K_2}) \rbrace\rbrace$} (or3)
  (or2) edge [bend right=10] node[pos=0.75,right=5pt] {\scriptsize 5. $\lbrace\lbrace g^{y_2}, H(K_2|hs_{K_2}) \rbrace\rbrace$} (or1)
  (or1) edge [bend right=10] node[pos=0.15,right] {\scriptsize 6. $\lbrace\lbrace g^{y_2}, H(K_2|hs_{K_2}) \rbrace\rbrace$} (client);

\end{tikzpicture}

%% file: fairtor.bbl
\begin{thebibliography}{10}

\bibitem{af96}
M.~Abe and E.~Fujisaki.
\newblock How to date blind signatures.
\newblock In {\em ASIACRYPT}, pages 244--251, 1996.

\bibitem{atk11}
M.~H. Au, P.~P. Tsang, and A.~Kapadia.
\newblock {PEREA}: Practical {TTP}-free revocation of repeatedly misbehaving
  anonymous users.
\newblock {\em ACM Trans. Inf. Syst. Secur.}, 14(4):29, 2011.

\bibitem{bcly07}
V.~Benjumea, S.~G. Choi, J.~Lopez, and M.~Yung.
\newblock Anonymity 2.0 - {X}.509 extensions supporting privacy-friendly
  authentication.
\newblock In {\em CANS}, pages 265--281, 2007.

\bibitem{bcly08}
V.~Benjumea, S.~G. Choi, J.~Lopez, and M.~Yung.
\newblock Fair traceable multi-group signatures.
\newblock In {\em Financial Cryptography}, pages 231--246, 2008.

\bibitem{bfpv13}
O.~Blazy, G.~Fuchsbauer, D.~Pointcheval, and D.~Vergnaud.
\newblock Short blind signatures.
\newblock {\em Journal of Computer Security}, 21(5):627--661, 2013.

\bibitem{brands93}
S.~Brands.
\newblock Untraceable off-line cash in wallets with observers (extended
  abstract).
\newblock In {\em CRYPTO}, pages 302--318, 1993.

\bibitem{bl12}
E.~Brickell and J.~Li.
\newblock {E}nhanced {P}rivacy {ID}: A direct anonymous attestation scheme with
  enhanced revocation capabilities.
\newblock {\em Dependable and Secure Computing, IEEE Transactions on},
  9(3):345--360, 2012.

\bibitem{abc4trust}
J.~Camenisch, I.~Krontiris, A.~Lehmann, G.~Neven, C.~Paquin, K.~Rannenberg, and
  H.~Zwingelberg.
\newblock D2.1 architecture for attribute-based credential technologies -
  version 1, 2011.

\bibitem{cl02}
J.~Camenisch and A.~Lysyanskaya.
\newblock Dynamic accumulators and application to efficient revocation of
  anonymous credentials.
\newblock In {\em CRYPTO}, pages 61--76, 2002.

\bibitem{cmk10}
A.~Chaabane, P.~Manils, and M.~A. K{\^a}afar.
\newblock Digging into anonymous traffic: A deep analysis of the tor
  anonymizing network.
\newblock In {\em NSS}, pages 167--174, 2010.

\bibitem{chaum82}
D.~Chaum.
\newblock Blind signatures for untraceable payments.
\newblock In {\em CRYPTO}, pages 199--203, 1982.

\bibitem{ch91}
D.~Chaum and E.~van Heyst.
\newblock Group signatures.
\newblock In {\em EUROCRYPT}, pages 257--265, 1991.

\bibitem{czms06}
X.~Chen, F.~Zhang, Y.~Mu, and W.~Susilo.
\newblock Efficient provably secure restrictive partially blind signatures from
  bilinear pairings.
\newblock In {\em Financial Cryptography}, pages 251--265, 2006.

\bibitem{dp07}
C.~Diaz and B.~Preneel.
\newblock Accountable anonymous communication.
\newblock In {\em Security, Privacy, and Trust in Modern Data Management},
  Data-Centric Systems and Applications, pages 239--253. Springer Berlin
  Heidelberg, 2007.

\bibitem{dar14}
J.~Diaz, D.~Arroyo, and F.~B. Rodriguez.
\newblock New x.509-based mechanisms for fair anonymity management.
\newblock {\em Computers {\&} Security}, 46:111--125, 2014.

\bibitem{dar15}
J.~Diaz, D.~Arroyo, and F.~B. Rodriguez.
\newblock \texttt{libgroupsig}: an extensible c library for group signatures.
\newblock {\em submitted}, 2015.

\bibitem{dms04}
R.~Dingledine, N.~Mathewson, and P.~F. Syverson.
\newblock Tor: The second-generation onion router.
\newblock In {\em {USENIX} Security Symposium}, pages 303--320, 2004.

\bibitem{douceur02}
J.~R. Douceur.
\newblock The sybil attack.
\newblock In {\em Peer-to-Peer Systems, First International Workshop, {IPTPS}
  2002, Cambridge, MA, USA, March 7-8, 2002, Revised Papers}, pages 251--260,
  2002.

\bibitem{dn92}
C.~Dwork and M.~Naor.
\newblock Pricing via processing or combatting junk mail.
\newblock In {\em Advances in Cryptology - {CRYPTO} '92, 12th Annual
  International Cryptology Conference, Santa Barbara, California, USA, August
  16-20, 1992, Proceedings}, pages 139--147, 1992.

\bibitem{isoiec200081}
{ISO}/{IEC}.
\newblock {ISO}/{IEC} {CD} 20008-1.2: Information technology - security
  techniques - anonymous digital signatures - part 1: General, 2012.

\bibitem{jl99}
W.-S. Juang and C.-L. Lei.
\newblock Partially blind threshold signatures based on discrete logarithm.
\newblock {\em Computer Communications}, 22(1):73--86, 1999.

\bibitem{kty04}
A.~Kiayias, Y.~Tsiounis, and M.~Yung.
\newblock Traceable signatures.
\newblock In {\em EUROCRYPT}, pages 571--589, 2004.

\bibitem{lpy12b}
B.~Libert, T.~Peters, and M.~Yung.
\newblock Group signatures with almost-for-free revocation.
\newblock In {\em CRYPTO}, pages 571--589, 2012.

\bibitem{mbgks08}
D.~McCoy, K.~S. Bauer, D.~Grunwald, T.~Kohno, and D.~C. Sicker.
\newblock Shining light in dark places: Understanding the {T}or network.
\newblock In {\em Privacy Enhancing Technologies}, pages 63--76, 2008.

\bibitem{okamoto06}
T.~Okamoto.
\newblock Efficient blind and partially blind signatures without random
  oracles.
\newblock In {\em TCC}, pages 80--99, 2006.

\bibitem{spc95}
M.~Stadler, J.-M. Piveteau, and J.~Camenisch.
\newblock Fair blind signatures.
\newblock In {\em EUROCRYPT}, pages 209--219, 1995.

\bibitem{tfs04}
I.~Teranishi, J.~Furukawa, and K.~Sako.
\newblock k-times anonymous authentication (extended abstract).
\newblock In {\em ASIACRYPT}, pages 308--322, 2004.

\bibitem{TPM03}
{T}rusted~{C}omputing {G}roup.
\newblock {TCG} {TPM} {S}pecification 1.2.
\newblock \url{http://www.trustedcomputinggroup.org}, 2003.

\bibitem{taks07}
P.~P. Tsang, M.~H. Au, A.~Kapadia, and S.~W. Smith.
\newblock Blacklistable anonymous credentials: blocking misbehaving users
  without ttps.
\newblock In {\em ACM Conference on Computer and Communications Security},
  pages 72--81, 2007.

\bibitem{tkcs11}
P.~P. Tsang, A.~Kapadia, C.~Cornelius, and S.~W. Smith.
\newblock Nymble: Blocking misbehaving users in anonymizing networks.
\newblock {\em IEEE Trans. Dependable Sec. Comput.}, 8(2):256--269, 2011.

\end{thebibliography}
